\documentclass[a4paper,10pt,twoside]{cpc-hepnp}
\usepackage{subfigure}
\usepackage{mathrsfs}
\usepackage{multicol}
\usepackage{graphicx}
\usepackage{booktabs}
\usepackage{amssymb,bm,mathrsfs,bbm,amscd}
\usepackage[tbtags]{amsmath}
\usepackage{lastpage}
\usepackage{amsmath, amssymb}
\usepackage{slashed}
\usepackage{multirow}
\usepackage{cepcphysics}
\usepackage{CJK}
\usepackage{lineno}
\usepackage{color}
\newcommand{\tabincell}[2]{\begin{tabular}{@{}#1@{}}#2\end{tabular}}
\newcommand{\lumiunit}{\ensuremath{\mathrm{cm}^{-2}\mathrm{s}^{-1}}}
\newcommand{\ten           }[2]{\ensuremath{{#1}\times 10^{#2}}}
\newcommand{\percent}[1]{{#1}\,\%}

\newcommand{\llh          }{\ensuremath{l^{+}l^{-}H}}
\newcommand{\eeh          }{\ensuremath{e^{+}e^{-}H}}
\newcommand{\mmh          }{\ensuremath{\mu^{+}\mu^{-}H}}

\newcommand{\bpair           }{\ensuremath{b\bar{b}}}
\newcommand{\cpair           }{\ensuremath{c\bar{c}}}
\newcommand{\gpair           }{\ensuremath{gg}}
\newcommand{\qpair           }{\ensuremath{q\bar{q}}}

\newcommand{\elpair           }{\ensuremath{e^{+}e^{-}}}

\newcommand{\mupair           }{\ensuremath{\mu^{+}\mu^{-}}}

\newcommand{\taupair           }{\ensuremath{\tau^{+}\tau^{-}}}

\newcommand{\mum               }{\ensuremath{\mathrm{\mu m}}}
\definecolor{orange}{rgb}{1,0.5,0}
\definecolor{violet}{rgb}{0.5,0,0.5}

\begin{document}
\begin{CJK*}{GBK}{song}

\fancyhead[c]{\small Chinese Physics C~~~Vol. xx, No. x (2019)
xxxxxx} \fancyfoot[C]{\small xxxxx-\thepage}

\footnotetext[0]{Received \today}

\title{Measurements of decay branching fractions of $H\to\bpair/\cpair/\gpair$ in associated $(\elpair/\mupair)H$ production at the CEPC\thanks{Supported by the National Key Program for S\&T Research and Development~(2016YFA0400400), by the National Natural Science Foundation of China (NSFC) under Grants~(11705028), by the Fundamental Research Funds for the Central Universities~(2242015K40027 and 2242017K40018), by the Provincial Innovation and Entrepreneurship Training Program for Undergraduate~(201710286068X), and by the Beijing Municipal Science and Technology Commission project~(Z181100004218003).}}

\author{%
Yu Bai$^{1;1)}$\email{baiy@seu.edu.cn}%
\quad Chun-Hui Chen$^{2;2)}$\email{cchen23@iastate.edu}%
\quad Ya-Quan Fang$^{3;3)}$\email{fangyq@ihep.ac.cn}%
\quad Gang Li$^{3;4)}$\email{li.gang@mail.ihep.ac.cn}
\quad Man-Qi Ruan$^{3;5)}$\email{manqi.ruan@ihep.ac.cn}\\
\quad Jing-Yuan Shi$^{1,4}$
\quad Bo Wang$^{1,5}$
\quad Pan-Yu Kong$^{1,6}$
\quad Bo-Yang Lan$^1$
\quad Zhan-Feng Liu$^1$
}
\maketitle

\address{%
$^1$ School of Physics, Southeast University, Nanjing, 210096, China\\
$^2$ Department of Physics and Astronomy, Iowa State University, Ames 50011-3160, USA\\
$^3$ Institute of High Energy Physics, Chinese Academy of Sciences, Beijing 100049, China\\
$^4$ School of physics and astronomy, the University of Manchester, Oxford Rd, M13 9PL, UK\\
$^5$ Department of Physics, College of Sciences, Northeastern University, Shenyang 110004, China\\
$^6$ School of Cyber Science and Engineering, Southeast University, Nanjing, 210096, China
}

\begin{abstract}
The high-precision measurement of Higgs boson properties is one of the primary goals of the Circular Electron Positron Collider (CEPC).
The measurements of $H\to\bpair/\cpair/\gpair$ decay branching fraction in the CEPC experiment is presented, considering a scenario of analysing 5000~\ifb~\elpair~collision data with the center-of-mass energy of 250~\GeV.
In this study the Higgs bosons are produced in association with a pair of leptons, dominantly mediated by the $ZH$ production process.
The statistical uncertainty of the signal cross section is estimated to be about 1~\% in
the $H\to b\bar{b}$ final state, and approximately 5-10~\% in the
$H\to c\bar{c}/gg$ final states. In addition, the main sources of the systematic uncertainties
and their impacts to the measurements of branching fractions are discussed.
This study demonstrates the potential of precise measurement of the hadronic final states of the Higgs boson decay at the CEPC, and will provide key information to understand the Yukawa couplings between the Higgs boson and quarks, which are predicted to be the origin of quarks' masses in the Standard Model.
\end{abstract}



\footnotetext[0]{\hspace*{-3mm}\raisebox{0.3ex}{$\scriptstyle\copyright$}2013
Chinese Physical Society and the Institute of High Energy Physics
of the Chinese Academy of Sciences and the Institute
of Modern Physics of the Chinese Academy of Sciences and IOP Publishing Ltd}%

\begin{multicols}{2}

\section{Introduction}

The discovery of a scalar boson with a mass around 125$\GeV$ at the Large Hadron Collider (LHC)~\cite{Higgs_ATLAS, Higgs_CMS} completed the final piece of the standard model (SM).
This particle, interpreted as the Higgs boson, plays a crucial role in the Electroweak Spontaneous Symmetry Breaking (EWSB), known as the Higgs mechanism~\cite{BEH, BEH2, BEH3}.
The Higgs mechanism allows the $\Wboson$ boson and $\Zboson$ boson to be massive while keeping the $SU(2)_L \times U(1)_Y$ gauge invariance.~As a consequence of this mechanism, the fermions like quarks and charged leptons get their masses from their couplings to the Higgs field.
The masses of the fermions ($m_{f_i}$) in the SM are proportional to their Yukawa couplings ($h_i$) to the Higgs field: $m_{f_i} = vh_i/\sqrt{2}$, where $v\approx 246\GeV$ is the vacuum expectation value of the Higgs field.
Thus measuring the Yukawa couplings between the Higgs boson and the SM fermions is essential to understand the origin of the fermions' masses.
The deviation of these couplings from SM prediction would indicate new physics.
\par
The dominant Higgs boson decays into fermionic final states are $\Hboson\to \bpair$, $\Hboson\to\taupair$ and $\Hboson\to\cpair$,
the decay branching fractions of which are predicted to be \percent{57}, \percent{6} and \percent{2.7} respectively~\cite{Higgs_qq_1,Higgs_qq_2}.
In addition, the Higgs boson can decay to a pair of gluons via heavy quark loops.~The large coupling between the Higgs boson and the top quark leads to considerably large
branching fraction of $\Hboson\to\gpair$ which is estimated to be about \percent{9}~\cite{Higgs_gg_1,Higgs_gg_2}.
\par
Until now, all the Higgs boson measurements are performed in hadron colliders.~The leading fermionic Higgs boson decay, $\Hboson\to\bpair$, was studied in both ATLAS and CMS, using the LHC Run-I data, which contains about 5~\ifb and 20~\ifb~of $pp$ collision data with $\sqrt{s}$ of 7~\TeV~and 8~\TeV~respectively.~These measurements include several Higgs boson production channels: $VH$~\cite{VH_bb_atlas,VH_bb_cms}, $t\bar{t}H$~\cite{ttH_bb_cms_1,ttH_bb_cms_2,ttH_bb_atlas_1} and VBF~\cite{VBF_bb_atlas,VBF_bb_cms} processes.
The $\Hboson\to\bpair$ were also studied at Tevatron~\cite{Tevatron_VH_bb} in $VH$ production, using 9.7~\ifb $p\bar{p}$ collision data with $\sqrt{s}$ of 1.96~\TeV.
The $H\to\bpair$ signal strength, defined as the ratio of the measured cross section to the corresponding SM prediction, is estimated to be $0.70\pm0.29$ according to the combination of ATLAS and CMS analysis of run-I data~\cite{higgs_atlas_cms_combine}.
In 2018 observations of $\Hboson\to\bpair$ decay in $VH$ production were declared by ATLAS~\cite{atlas_vhbb_new} and CMS~\cite{cms_vhbb_new}, using 79.8~\ifb and 41.3~\ifb~ $pp$ collision data with $\sqrt{s}$ of 13~\TeV.
The signal strength is $1.16\pm 0.16(stat.)^{+0.21}_{-0.19}(sys.)$ and $1.01\pm0.22$ respectively.
The $\Hboson\to\cpair$ was also studied using 36.1~\ifb data with $\sqrt{s}$ of 13~\TeV~in ATLAS~\cite{H_cc_atlas}, giving a upper limit about 100 times higher than the SM prediction with 95\% confidence level.
The precision of those results is limited by large QCD production background, which is inevitable in hadron colliders. \par
A lepton collider has significant advantage in precise Higgs measurements as it's free of QCD production background and a has precise and tunable initial energy.
Several future lepton colliders have been proposed with the capability of precise measurement of Higgs boson parameters, such as the International Linear Collider (ILC)~\cite{ILC}, the $\elpair$ Future Circular Collider (FCC-ee)~{\cite{TLEP}}, the Compact Linear
Collider (CLIC)~\cite{CLIC} and the Circular Electron Positron Collider (CEPC)~\cite{CEPC_preCDR}. The CEPC is a proposed electron-positron collider by the Chinese high energy physics community.~It can be operated with $\sqrt{s}$ of 240~\GeV~to 250~\GeV~and the designed instaneous luminosity is \ten{2}{34} \lumiunit.
The cross section of Higgs production is about 0.2 $\mathrm{pb}$ in CEPC with $\sqrt{s}$ of 250~\GeV.
The primary production process is via $ZH$ production(96.6\%), which is often referred as Higgs-strahlung~\cite{Higgs_strahlung_1,Higgs_strahlung_2,Higgs_strahlung_3} process,
while the fraction of production via $WW$-fusion~\cite{WW_fusion} and $ZZ$-fusion is $3.06\,\%$ and $0.29\,\%$ respectively.
After ten years of running, one million of the Higgs boson (5000 $\ifb$ collision data) are expected to be collected
at the CEPC.\par
The work presented here focuses on the Higgs production in association with a pair of leptons~($\elpair$ or $\mupair$), in which Higgs decays to a pair of $b$-quark, $c-$quark or gluons. The leptons are either from $Z-$boson decay in Higgs-strahlung process, or $ZZ-$fusion which takes place only in $\eeh$ channel.
The measurements of signal cross sections, denoted as $\sigma^{H\to\bpair}_{\llh}$, $\sigma^{H\to\cpair}_{\llh}$ and $\sigma^{H\to\gpair}_{\llh}$, are described.
The branching fraction of $H\to\bpair/\cpair/\gpair$ can be derived once the cross sections of the $\llh$ production, $\sigma_{\eeh}$ and $\sigma_{\mmh}$, are determined from other measurements.

The work presented in this paper is partly inspired by the $H\to\bpair/\cpair/\gpair$ analysis in ILC~\cite{ILC:Hbbccgg}, and it is the subsequent study in $H\to\bpair/\cpair/\gpair$ analysis presented in CEPC Higgs white paper~\cite{CEPC_Higgs} with improvement in background estimation.
\par
This paper is organized as follows.~After the introduction, a brief description of the detector will be presented in {Sec.~{\ref{sec:CEPC}}.
The MC samples and event selections are described in Sec.~{\ref{sec:MC_Sample}} and Sec.~{\ref{sec:event_selection}} respectively.
In Sec.~{\ref{sec:templatefit}}, the flavor tagging and flavor-template-recoil-mass-fit, which is the procedure to extract signal event yields, is presented.
In Sec.~{\ref{sec:uncertainties}}, the uncertainties of the signal cross sections and signal branching fractions are discussed, and finally in Sec.~{\ref{sec:summary}} a short summary is provided.}\par

\section{Detector Design}
\label{sec:CEPC}
 The detailed description of the proposed CEPC detector can be found elsewhere~\cite{CEPC_preCDR}.
A vertex detector with high pixel resolution is located in the inner
most part of the detector.
The 6 layers of sensors are laid coaxially, in radius from 16~mm to 60~mm, covering 97\% - 90\% in the range of the polar angle.
The spatial resolution in a single layer is 2.8 $\mu$m in the 2 inner
layers and 4 $\mu$m in the 4 outer layers. The overall IP resolution can be represented as:
\begin{equation}
{\sigma(r\phi)}= a \oplus \frac{b}{p(\mathrm{\GeV}) \sin^{3/2} \theta },\mum
\end{equation}
The first and second term in the right side of equation depict the resolution from finite single point position and the resolution due to multi-scattering respectively.
These two types of resolution are parameterized as $a$ and $b$, which are estimated to be 5 and 10 respectively in CEPC.
The parameters $p$ and $\theta$ are the momentum and polar angle of the reconstructed charged particle and $\oplus$ denotes summation in quadrature.
The high spatial resolution is essential to track impact parameter
 (IP) measurements and vertex reconstruction, on which the identification of the flavor of jets (flavor tagging) primarily relies.
\par 
A Time Project Chamber~(TPC) is located outside of the vertex dector to take the major task of track
measurement. It covers the solid angle up to $\cos \theta = 0.98$.
When being operated in designed magnetic field of 3.5~T, the momentum resolution is
$\sigma(1/p_T) = 10^{-4}\GeV$. \par
The calorimeter system includes two sub-systems: the electromagnetic calorimeter~(ECAL) and the hadronic calorimeter~(HCAL).
They are designed to have high energy resolution as well as high spatial resolution.
The ECAL is a silicon-tungsten-based detector, which uses tungsten as absorber and silicon as sensor. It contains 30 layers of sampling structures. Each layer is divided into cells of $5\mathrm{mm}\times 5\mathrm{mm}$ in size.
The HCAL includes 40 layers. Each layer contains a 20 mm thick stainless steel as absorber layer, a 3 mm thick glass resistive plate chamber~(GRPC) and a 3 mm thick readout electronics in $1\mathrm{cm}\times 1\mathrm{cm}$ readout pad.
The overall jet energy resolution(JER) is 3-4\%, and the two-jet system invariant mass resolution is required to be around 3-4\%, to distinguish the final states from $Z$ boson and $W^{\pm}$ boson hadronic decay. The high granularity of the ECAL and HCAL is crucial for application of the particle flow algorithm~\cite{PFA}, which intends to reconstruct and identify each particle individually by combining information from all sub-detectors.
The details of the ECAL and HCAL design and performance can be found in Ref.~\cite{CEPC_CDR}.
\par
The muon system is mounted at the outermost part of the detector. The baseline
design of muon detector has 8 sensitive layers in barrel and endcap region.
On average, muons with momentum 2~\GeV~are expected to hit the first layer while those with momentum over 4~\GeV~can penetrate all the 8 layers.~Muons with momentum above 5~\GeV~can be detected with standalone muon identification efficiency above 95\% while the fake rate of pions is less than 1\%.~In order to improve the precision of the muon momentum measurement, the longitude and transverse position resolution are required to be $\sigma _{z} = 1.5 \cm$ and $\sigma_{r\phi} = 2.0 \cm$ respectively.
\par

\section{MC Sample}~\label{sec:MC_Sample}
Both background and signal events are generated using Whizard~\cite{Wizard_1} configured as no-polarization electron-positron collision with the $\sqrt{s}$ of 250~\GeV. The mass of the Higgs boson is assumed to be 125~\GeV~and the couplings are set to the SM predictions.
The fragmentation and hadronization are performed by implementing PYTHIA{6}~\cite{PYTHIA64}.
All the MC datasets are normalized to the expected yields in data with integrated luminosity of 5000~\ifb, by assigning a weight to the events of each process.~The details of the event generation in CEPC can be found at Ref.~\cite{CEPC_Generation}.
\par
The generated events undergo the detector simulation by Mokka~\cite{mokka}, a GEANT4~\cite{Geant4} based detector simulator. The simulated hits are digitized and reconstructed with ArborPFA~\cite{ArborPFA_1,ArborPFA_2,CEPC:reconstruction}.\par
The charged particles are identified as electrons or muons by Lepton Identification in Calorimeter with High Granularity(LICH) algorithm, a dedicate lepton identification algorithm designed for Higgs factories, as described in Ref.~\cite{LICH}. The algorithm use the $dE/dX$ information measured by TPC, together with shower and hit information in high granularity calorimeter, as discrimination variables. A Boosted Decision Trees algorithm~\cite{BDT} with Gradient boosting~(GBDT) method is implemented to further extract the discriminative characteristics of the variables.~The overall efficiencies for electron and muons are 99.7\% and 99.9\% respectively, with the rate of electron and muons misidentified as each other smaller than 0.07\%. The rate of particles like $\pi^{\pm}$ identified as electrons or muons is 0.21\% and 0.05\% respectively.
\par
Jets reconstruction and flavor tagging are essential to this analysis.
They are done with the LCFIPLUS~\cite{LCFIPlus} software package, integrating the functionality of vertex finding, jet reconstruction and jet flavor tagging.
Before the jet clustering, the secondary vertices are identified based on the reconstructed tracks.
Jets are reconstructed by Durham algorithm{~\cite{Durham}}.
This algorithm begins with jet cluster candidates, which are either single reconstructed particles, or compound objects like reconstructed secondary vertices.
The procedure iteratively pairs the clusters and calculates the distance between them, defined as $y_{ij} = \mathrm{min}\{E^2_i,E^2_j\}(1-\cos\theta_{ij})/E^2_{vis}$, where $E_i$ and $E_j$ are the energy of
$i$-th and $j$-th cluster, and $\theta_{ij}$ refers to the angle between them.
$E_{vis}$ are the sum of energy of all the clusters in the event.
Clusters with minimum $y_{ij}$ are merged, reducing the cluster number by 1, until the ceasing criteria are met.
The ceasing criteria can be either a minimum $y_{ij}-$value threshold, or the remaining clusters number equals to the required jet multiplicity.
In this analysis, the $y_{ij}$ threshold is set to 0, and each event is forced to have two jets reconstructed.
The minimum value $y_{ij}$ can be denoted as $Y_k$, in which $k$ is the multiplicity of cluster candidates.
When $k$ is larger than 2, the signal events have relatively smaller $Y_k$ than the backgrounds with at least $k$ primary parton\footnote{Here primary parton refers to partons produced directly via electroweak processes or Higgs boson decay. It only counts for parton multiplicity before gluon radiation and gluon splitting.}.
This is because in a signal event, the closet two clusters, among more of them, are likely to be from the same parton. They tend to be collinear, which lead to small $Y_k$.\par

\section{Event selection}\label{sec:event_selection}

The final state of the signal contains two jets and two leptons with opposite charge and same flavor.
There are two types of backgrounds according to the final states: the irreducible backgrounds and the reducible backgrounds.
The irreducible backgrounds contain the same final states as that in signal.~
The semi-leptonic $ZZ$ process, in which one $Z-$boson decays to $\elpair$ or $\mupair$ and the other decays to quark pair, is a typical example and the major components of the irreducible backgrounds.
The reducible backgrounds include all the other types of background which have different final states, such as hadronic or leptonic $W^+W^-$ and $ZZ$ production, the lepton pair or quark pair production, semi-leptonic $W^+W^-$ production, and the Higgs production processes other than $\llh$.
There are backgrounds from $\eeh$ or $\mmh$ production, in which the Higgs boson decays to $WW^*$ or $ZZ^*$ and subsequently both vector bosons decay to quarks.
They are reducible backgrounds since there are more quarks in final states than that in signal. But they behave more like irreducible background experimentally, so they will be discussed separately from the typical irreducible or reducible backgrounds, and they will be referred as $\eeh$ or $\mmh$ background in this analysis.
There are also other irreducible backgrounds from $\llh$, in which the Higgs boson decays to light quark pair, to $\tau^+\tau^-$ or to photons when all the tauons or photons from the Higgs boson decay are misidentified as jets. Their contributions are expected to be very small and they are classified as reducible backgrounds instead of $\llh$ backgrounds.
\par
Each event must contain two isolated tracks with opposite charge, reconstructed as $\elpair$ or $\mupair$.
The energy of each isolated lepton candidate must be above 20~\GeV.
The isolation criterion requires $E^2_{\mathrm{cone}}<4 E_{\mathrm{lep}}+12.2$, where $E_{\mathrm{lep}}$ is the energy of the lepton, and $E_{\mathrm{cone}}$ is the energy within a cone $\cos\theta_{\mathrm{cone}}>0.98$ around the lepton. Here $E_{\mathrm{lep}}$ and $E_{\mathrm{cone}}$ are measured in~\GeV.
Events with additional isolated leptons are rejected (extra isolated lepton veto).
The polar angle of lepton pair system is required to be in the range of $|\cos\theta_{\mupair}| < 0.81$ and $|\cos\theta_{\elpair}| < 0.71$. The
angle between the two isolation tracks $\psi$ is required to satisfy $\cos\psi > -0.93$ and $\cos\psi > -0.74 $ for $\elpair H$ and $\mupair H$ channel respectively, to reject events from lepton pair production where leptons tend to be back to back.
The invariant mass of lepton pair is required to be inside the $Z-$mass window, which is defined as 77.5~$\GeV$ to 104.5~$\GeV$.
\par
The remaining particles in the event are used to reconstruct exactly two jets with polar angle $\theta_{\rm jet}$ in the range of $|\cos\theta_{\rm jet}|<0.96$.
The two jets are required to contain at least 20 particles, each with energy no less than 0.4~\GeV, according to the optimization of distinguishing the signal events against the events including fake jets from photons or leptons.~
The invariant mass of the pair of jets is required to be between $75~\GeV$ and $150~\GeV$ to reject the irreducible backgrounds.

To suppress the $\eeh$ and the $\mmh$ backgrounds, $Y_4$, an indicator of primary parton multiplicity described in Sec~.{\ref{sec:MC_Sample}}, was required to be less than 0.011.\par
The invariant mass of the lepton pair's recoil system, denoted as $M^{l\bar{l}}_{\mathrm{recoil}}$, provides clear signature of the $l\bar{l}H$ events.
The definition of $M^{l\bar{l}}_{\mathrm{recoil}}$ is:
\begin{equation}\label{eqn:Mrecoil}
M^{l\bar{l}}_{\mathrm{recoil}} = \sqrt{(\sqrt{s}-E_l-E_{\bar{l}})^2-(\vec{P}_l+\vec{P}_{\bar{l}})\cdot(\vec{P}_l+\vec{P}_{\bar{l}})},
\end{equation}
in which $\sqrt{s} = 250$~\GeV, while $E$ and $\vec{P}$ stand for the energy and momentum of the leptons respectively.
A Higgs mass window is defined by requiring $M^{l\bar{l}}_{\mathrm{recoil}}$ between 124~\GeV and 140~\GeV.
The signal and background yields during the event selections are summarized in Tab.~\ref{tab:cutflow} for $\mmh$ and $\eeh$ analysis, respectively.
\par

\end{multicols}

\begin{center}
\tabcaption{ \label{tab:cutflow}  Event yields of cut flow. Signal events are $l\bar{l}+H\to l\bar{l}+\bpair/\cpair/\gpair$ combined. $\mmh$ and $\eeh$ background refers to the background which Higgs are produced associated with $\mupair$ and $\elpair$, but decay to final states other than $\bpair/\cpair/\gpair$. 'Other Higgs background' stands for the Higgs production process different from the signal. 'Irreducible SM background' is the $\elpair/\mupair+$jet pair process without Higgs productions. 'Other SM background' includes all the other background processes. The 'fit region' will be described in Sec.~\ref{sec:templatefit}}
\footnotesize
\begin{tabular*}{170mm}{@{\extracolsep{\fill}}cccccc}
\toprule \multicolumn{6}{c}{$\mmh\to \mupair+\bpair/\cpair/\gpair$ Channel}\\
\hline
          &   Signals &  \tabincell{c}{$\mmh$\\ Background} &\tabincell{c}{Other Higgs \\Background} &  \tabincell{c}{Irreducible\\ Background}& \tabincell{c}{Other SM\\Background} \\ \hline
  Original     &   \ten{2.45}{4}   &   \ten{1.10}{4}  &  \ten{1.01}{6} & \ten{1.05}{6} &  \ten{4.96}{8} \\ \hline
     Lepton pair selection without recoil mass cut &   \ten{1.51}{4}   &   \ten{6.56}{3}  &  227 & \ten{1.09}{4} &  \ten{2.79}{4}  \\ \hline
     \tabincell{c}{Jets pair selection and \\ lepton pair recoil mass cut for fit region} &   \ten{1.32}{4}   &   \ten{1.80}{3}  &  108 & \ten{7.75}{3} &  43.6 \\ \hline
    Signal Region & \ten{1.31}{4} & \ten{1.80}{3} & 96.1 & \ten{5.78}{3}   &  38.4    \\ \hline
     \multicolumn{6}{c}{$\eeh\to \elpair+\bpair/\cpair/\gpair$ Channel} \\ \hline
          &   Signals &  \tabincell{c}{$\eeh$\\ Background} &\tabincell{c}{Other Higgs \\Background} &  \tabincell{c}{Irreducible\\ Background}& \tabincell{c}{Other SM\\Background} \\ \hline
  Original     &   \ten{2.63}{4}   &   \ten{1.17}{4}  &  \ten{1.01}{6} & \ten{1.62}{6} &  \ten{4.95}{8} \\ \hline
     Lepton pair selection without recoil mass cut &   \ten{9.17}{3}   &   \ten{3.53}{3}  &  128 & \ten{9.00}{3} &  \ten{7.11}{4}  \\ \hline
     \tabincell{c}{Jets pair selection and \\recoil lepton pair mass cut of fit region} &   \ten{7.14}{3}   &   917  &  56.1 & \ten{8.63}{3} &  69.4\\ \hline
     Signal Region &  \ten{7.13}{3} &   913 &     36.4      & \ten{4.14}{3}      &    67.4  \\ \hline
\bottomrule
\end{tabular*}%
\end{center}

\begin{multicols}{2}
\section{Recoil-mass-flavor fit}{\label{sec:templatefit}}
After applying the object and event selection described in {Sec.~\ref{sec:event_selection}}, it is necessary to get the component fractions of $H\to\bpair$, $H\to\cpair$ and $H\to gg$ processes.
It is achieved by the high performance multi-variable-based flavor tagging toolkit in LCFIPLUS.
This toolkit is responsible for the flavor tagging algorithm training and implementing.
The training is applied to the simulated $Z\to q\bar{q}$ sample, produced with $\sqrt{s}$ of 91.2~\GeV.
The reconstructed jets in the sample are classified to 4 categories according to multiplicity the secondary vertex and lepton in the jet: jets with both secondary vertex and lepton, jets with secondary vertex but without lepton, jets without secondary vertex but with lepton and jets without neither secondary vertex nor lepton. The details of the secondary vertex finding can be find in Ref.~\cite{LCFIPlus}. The leptons are selected according to lepton identification presented in {Sec.\ref{sec:MC_Sample}} without any isolation criteria required.
In each category, two types of training, one for the $b$-tagging algorithm and the other for the $c$-tagging algorithm, are implemented with GBDT method.
Discrimination variables such as jets kinematic variables, impact parameters of tracks and secondary vertex parameters are included in the training.
After the training, a $b$-tagging model and a $c$-tagging model are created.
By invoking these models, a $b$-jet likeness weight and a $c$-jet likeness weight are calculated for each jet, representing the resemblance of the jet to a $b$-jet or a $c$-jet respectively. The likeness weights are in the range between 0 and 1, and a higher weight indicates higher likelihood of a jet to be $b-$jet or $c-$jet. \par
The $b$-weight likeness of the two individual jets of any selected event,
$L_{b1}$ and $L_{b2}$, can be used to construct the combined $B$ likeness, defined as:
 \begin{equation}
 \label{for:X_B_C}
  X_{B} = {L_{b1}L_{b2}}/[{L_{b1}L_{b2}+(1-L_{b1})(1-L_{b2})}].
  \end{equation}
  A combined $C-$likeness $X_C$ can be defined in similar way, by replacing the $L_{b1}$ and $L_{b2}$, by $L_{c1}$ and $L_{c2}$ respectively, in the right side of {Eq.~(\ref{for:X_B_C})}.
  The conservation of quark flavor in the Higgs boson decay guarantee that $X_B$($X_C$) is close to 1 if the Higgs boson decay to $b\bar{b}$~($c\bar{c}$) while close to 0 otherwise.
Thus the flavor of the jets in each event can be characterized by the two dimensional distribution of variable $X_B$ and $X_C$. Flavor templates are created for the $(X_B,X_C)$ distributions of different processes. By fitting the data with these flavor templates, one can get the fraction of each process. This template fit approach was implemented in ILC
$H\to\bpair/\cpair/\gpair$ analysis~\cite{ILC:Hbbccgg}.\par
\par



%
Here the template fit is combined with the fit to $M^{l\bar{l}}_{\mathrm{recoil}}$, defined in Eq.~(\ref{eqn:Mrecoil}).
Such a combined fit is motivated by separating the irreducible backgrounds from signal while extracting the flavor components in signal events.
The fit is applied to three variables: $M^{l\bar{l}}_{\mathrm{recoil}}$, $X_B$ and $X_C$ and simultaneously.
The RooFit package is implemented to perform an unbinned likelihood fit to the weighted events.
The overall likelihood function is constructed as:
\end{multicols}
\begin{eqnarray}
&&L(M^{l\bar{l}}_{\mathrm{recoil}},X_B,X_C; \vec{\theta_s},\vec{\theta_b},N^{\mathrm{sig}}_{H\to\bpair},N^{\mathrm{sig}}_{H\to\cpair},N^{\mathrm{sig}}_{H\to\gpair},N^{\mathrm{bkg}}_{\mathrm{irred}\_\bpair},N^{\mathrm{bkg}}_{\mathrm{irred}\_\cpair},N^{\mathrm{bkg}}_{\mathrm{irred}\_uds},N^{\mathrm{bkg}}_{\llh},N_{\mathrm{redu}})\nonumber\\
&=&P_{\mathrm{sig}}(M^{l\bar{l}}_{\mathrm{recoil}};\vec{\theta_s})(N^{\mathrm{sig}}_{H\to\bpair}P^{H\to\bpair}_{\mathrm{flavor}}(X_B,X_C)+N^{\mathrm{sig}}_{H\to\cpair}P^{H\to\cpair}_{\mathrm{flavor}}(X_B,X_C)+N^{\mathrm{sig}}_{H\to\gpair}P^{H\to\gpair}_{\mathrm{flavor}}(X_B,X_C)+N^{\mathrm{bkg}}_{\llh}P^{H\to\mathrm{other}}_{\mathrm{flavor}}(X_B,X_C))\nonumber\\
&+&P_{\mathrm{irred}}(M^{l\bar{l}}_{\mathrm{recoil}};\vec{\theta_b})(N^{\mathrm{bkg}}_{\mathrm{irred}\_\bpair}P^{\mathrm{irred}\_\bpair}_{\mathrm{flavor}}(X_B,X_C)+N^{\mathrm{bkg}}_{\mathrm{irred}\_\cpair}P^{\mathrm{irred}\_\cpair}_{\mathrm{flavor}}(X_B,X_C)+N^{\mathrm{bkg}}_{{\mathrm{irred}}\_uds}P^{\mathrm{irred}\_uds}_{\mathrm{flavor}}(X_B,X_C))\nonumber\\
&+&N_{\mathrm{redu}}P_{\mathrm{redu}}(M^{l\bar{l}}_{\mathrm{recoil}},X_B,X_C)
\label{for:Fit_Likelihood}
\end{eqnarray}
\begin{multicols}{2}

The $N^{\mathrm{sig}}_{H\to\bpair}$, $N^{\mathrm{sig}}_{H\to\cpair}$ and $N^{\mathrm{sig}}_{H\to\gpair}$ are the signal event yields of $H\to\bpair/\cpair/\gpair$ respectively, which are the concern of this analysis.
$N^{\mathrm{bkg}}_{\mathrm{irred}\_bb}$, $N^{\mathrm{bkg}}_{\mathrm{irred}\_cc}$ and $N^{\mathrm{bkg}}_{\mathrm{irred}\_uds}$ are the event yields of the irreducible background with $\bpair$, $\cpair$ or light quark in final states, respectively.
$N_{\mathrm{redu}}$ is the event yield of reducible backgrounds.
$\vec{\theta_s}$ and $\vec{\theta_b}$ are the shape parameters of $M^{l\bar{l}}_{\mathrm{recoil}}$ spectrum of $l\bar{l}H$ and irreducible background respectively.
Functions like $P^{\mathrm{p}}_{\mathrm{flavor}}(X_B,X_C)$ are the two-dimensional distributions of $(X_B,X_C)$ of process $p$, which are modeled as two-dimensional histograms with $20\times 20$ bins from MC simulation.
The recoil mass function of $\llh$ events, denoted as $P_{\mathrm{sig}}(M^{l\bar{l}}_{\mathrm{recoil}};\vec{\theta_s})$, are described by a crystal ball function, combined with a double sided exponential function which has peak near the Higgs mass threshold, to describe the resolution effect of track energy and momentum measurements.
The lepton pair recoil mass spectrum of irreducible background events, denoted as $P_{\mathrm{irred}}(M^{l\bar{l}}_{\mathrm{recoil}};\vec{\theta_b})$, is described by a first order Chebychev polynomial function~(\mmh~channel) or exponential functions~(\eeh ~ channel).
$P_{\mathrm{redu}}$ is the 3-dimensional distribution of $M^{l\bar{l}}_{\mathrm{recoil}}$, $X_B$ and $X_C$ of the reducible backgrounds. It is depicted by MC simulation.
The $M^{l\bar{l}}_{\mathrm{recoil}}$ ranges in the fit are set to be between 120~\GeV~and 140~\GeV~for~\mmh~channel, and between 115~\GeV and 140~\GeV~for~\eeh~channel. They are slightly wider than the signal regions to have better estimation of irreducible background. The signal and background yields in these regions are also summarized in Tab.~\ref{tab:cutflow}.
All the functions of $P_{\mathrm{sig}}(M^{l\bar{l}}_{\mathrm{recoil}};\vec{\theta_s})$, $P_{\mathrm{irred}}(M^{l\bar{l}}_{\mathrm{recoil}};\vec{\theta_b})$, $P^{\mathrm{p}}_{\mathrm{flavor}}(X_B,X_C)$ and $P_{\mathrm{redu}}(M^{l\bar{l}}_{\mathrm{recoil}},X_B,X_C)$ are normalized to 1 in the fit range.
\par

The shape parameters of the crystal ball function in $\vec{\theta_s}$ and all the parameters in $\vec{\theta_b}$ are free in the fit.
The tail distribution of signal events, in $M^{l\bar{l}}_{\mathrm{recoil}}$ spectrum near the Higgs mass threshold, are fixed according to a signal-only-fit.
The event yields parameters such as $N^{\mathrm{sig}}_{H\to\bpair}$, $N^{\mathrm{sig}}_{H\to\cpair}$, $N^{\mathrm{sig}}_{H\to\gpair}$, $N^{\mathrm{bkg}}_{\mathrm{irred}\_\bpair}$, $N^{\mathrm{bkg}}_{\mathrm{irred}\_\cpair}$ and $N^{\mathrm{bkg}}_{\mathrm{irred}\_uds}$ are also free in the fit, while $N^{\mathrm{bkg}}_{\llh}$ and $N_{\mathrm{redu}}$ are fixed to MC predictions. The 3D distribution of reducible background $P_{\mathrm{redu}}(M^{l\bar{l}}_{\mathrm{recoil}},X_B,X_C)$ is also fixed as that predicted by MC simulation.
\par
\begin{figure}
 \centering
 \subfigure[]
{
   \begin{minipage}[b]{0.47\textwidth}
   \includegraphics[width=\textwidth]{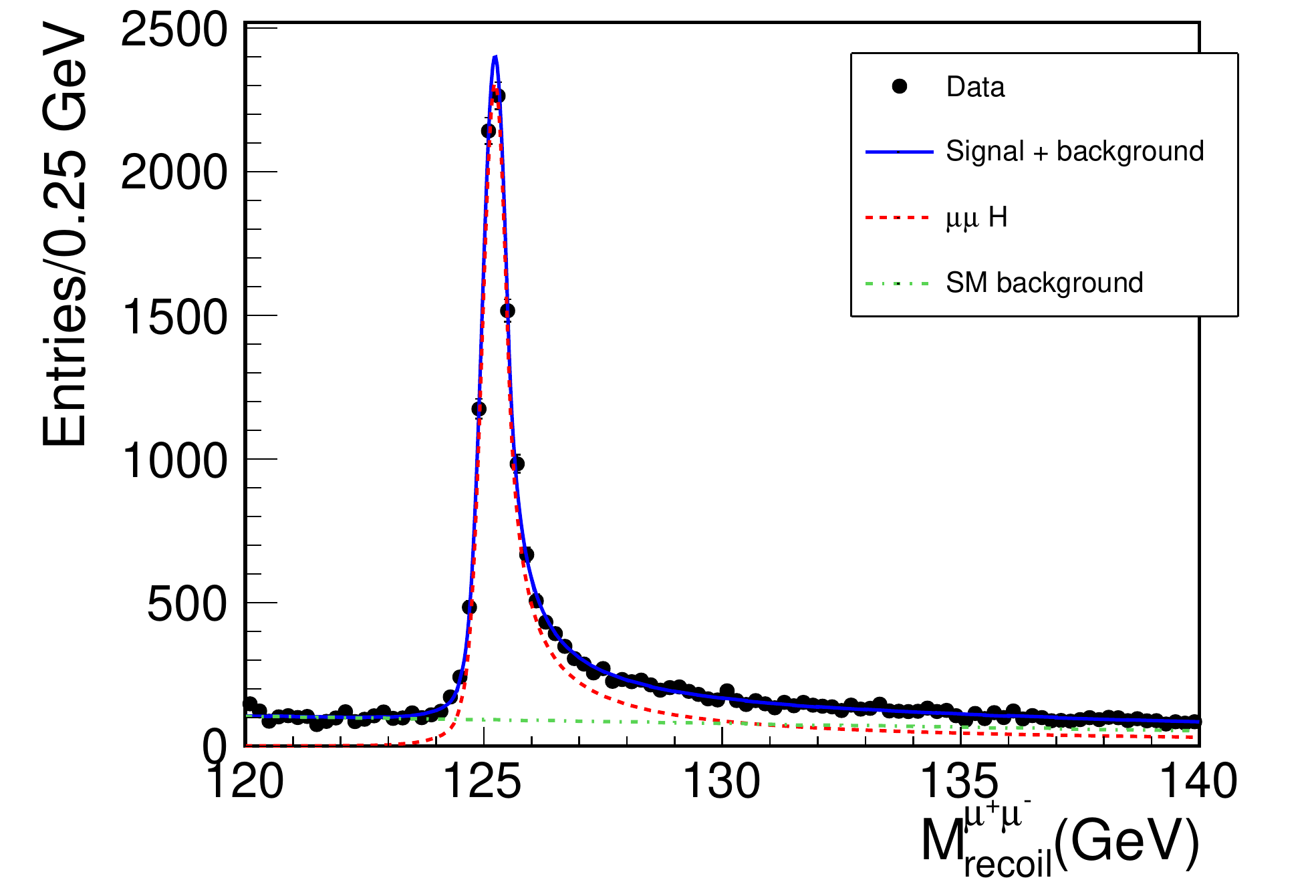}
   \end{minipage}
}
\subfigure[]
{
   \begin{minipage}[b]{0.47\textwidth}
   \includegraphics[width=\textwidth]{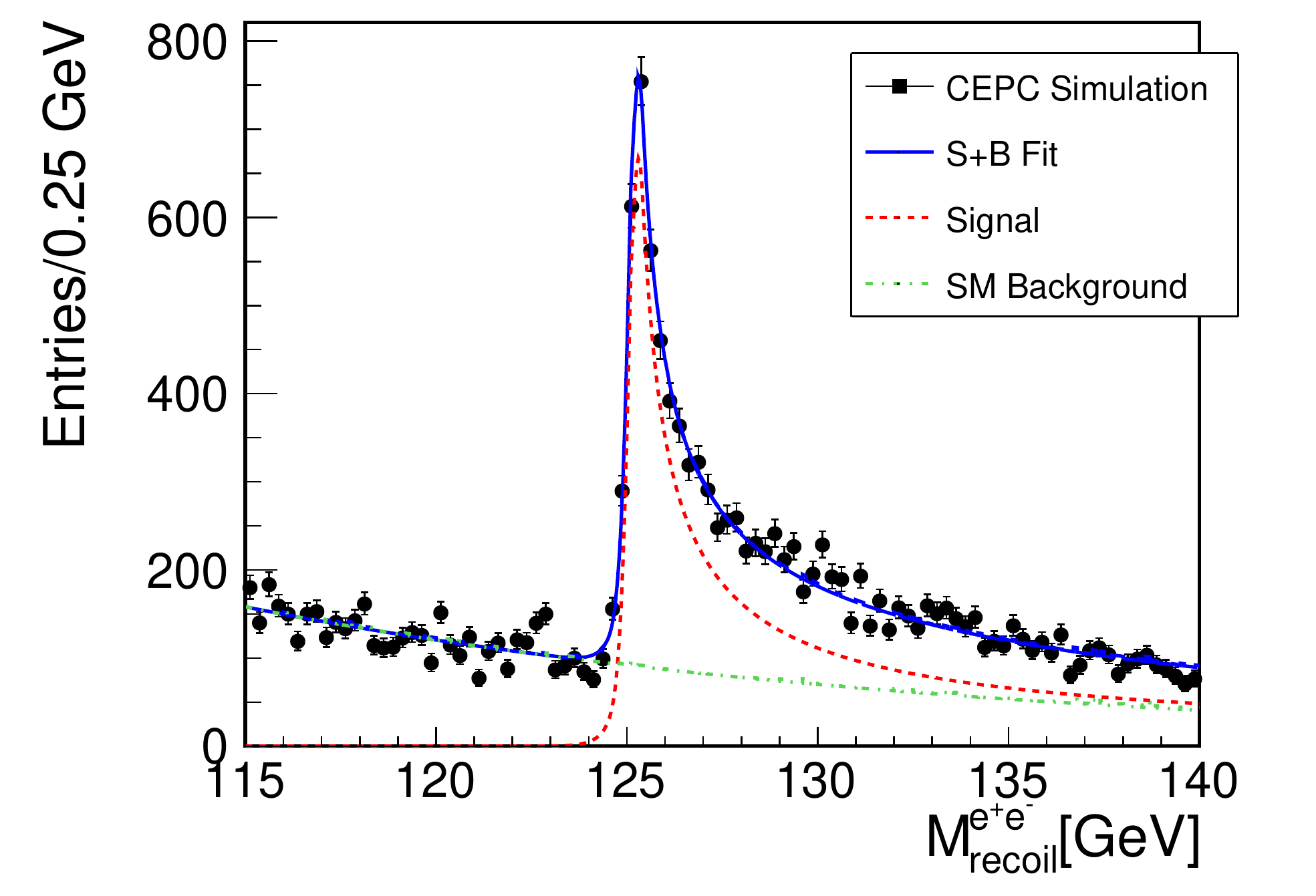}
   \end{minipage}
}
\subfigure[]
{
    \begin{minipage}[b]{0.47\textwidth}
    \includegraphics[width=\textwidth]{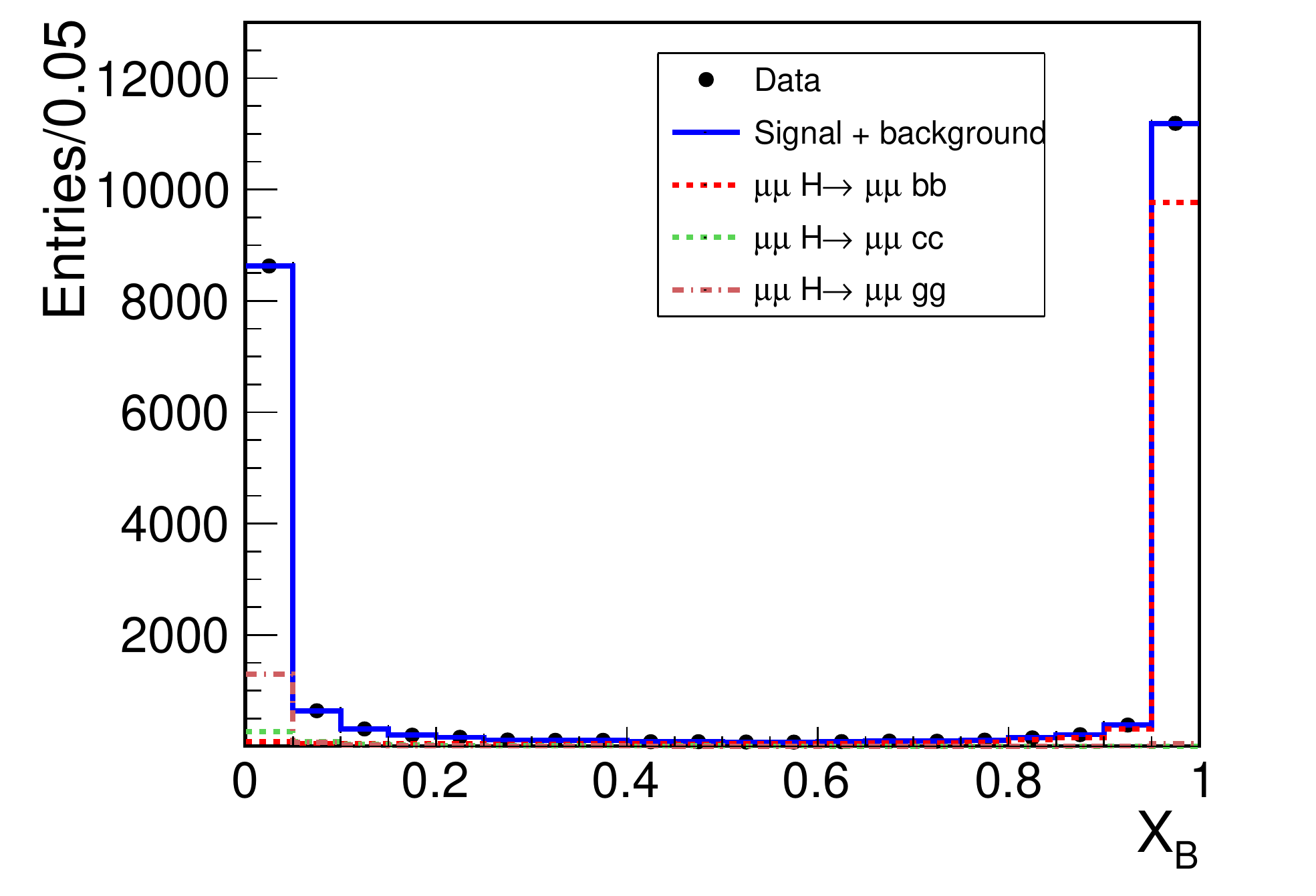}
    \end{minipage}
}
\subfigure[]
{
    \begin{minipage}[b]{0.47\textwidth}
    \includegraphics[width=\textwidth]{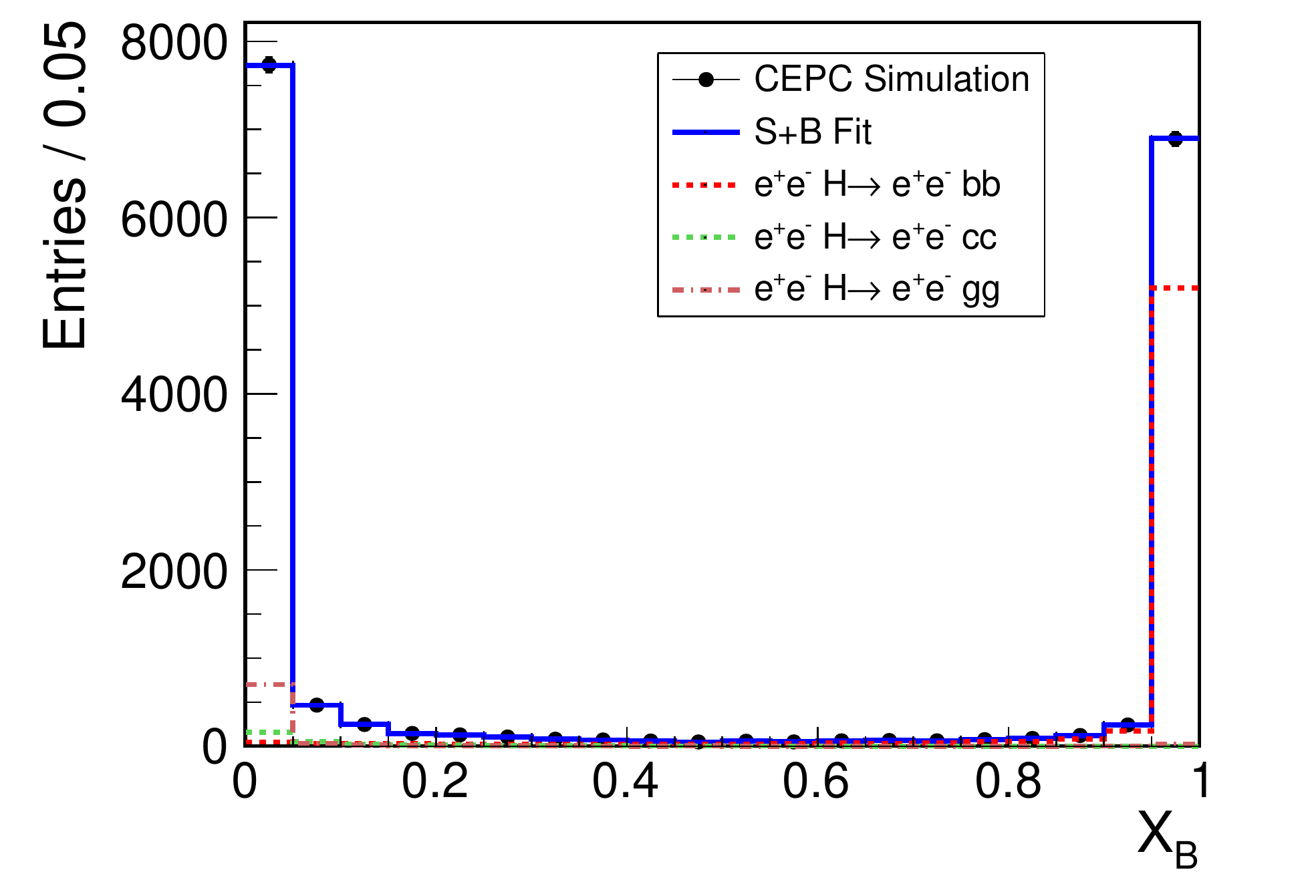}
    \end{minipage}
}
\subfigure[]
{
    \begin{minipage}[b]{0.47\textwidth}
    \includegraphics[width=\textwidth]{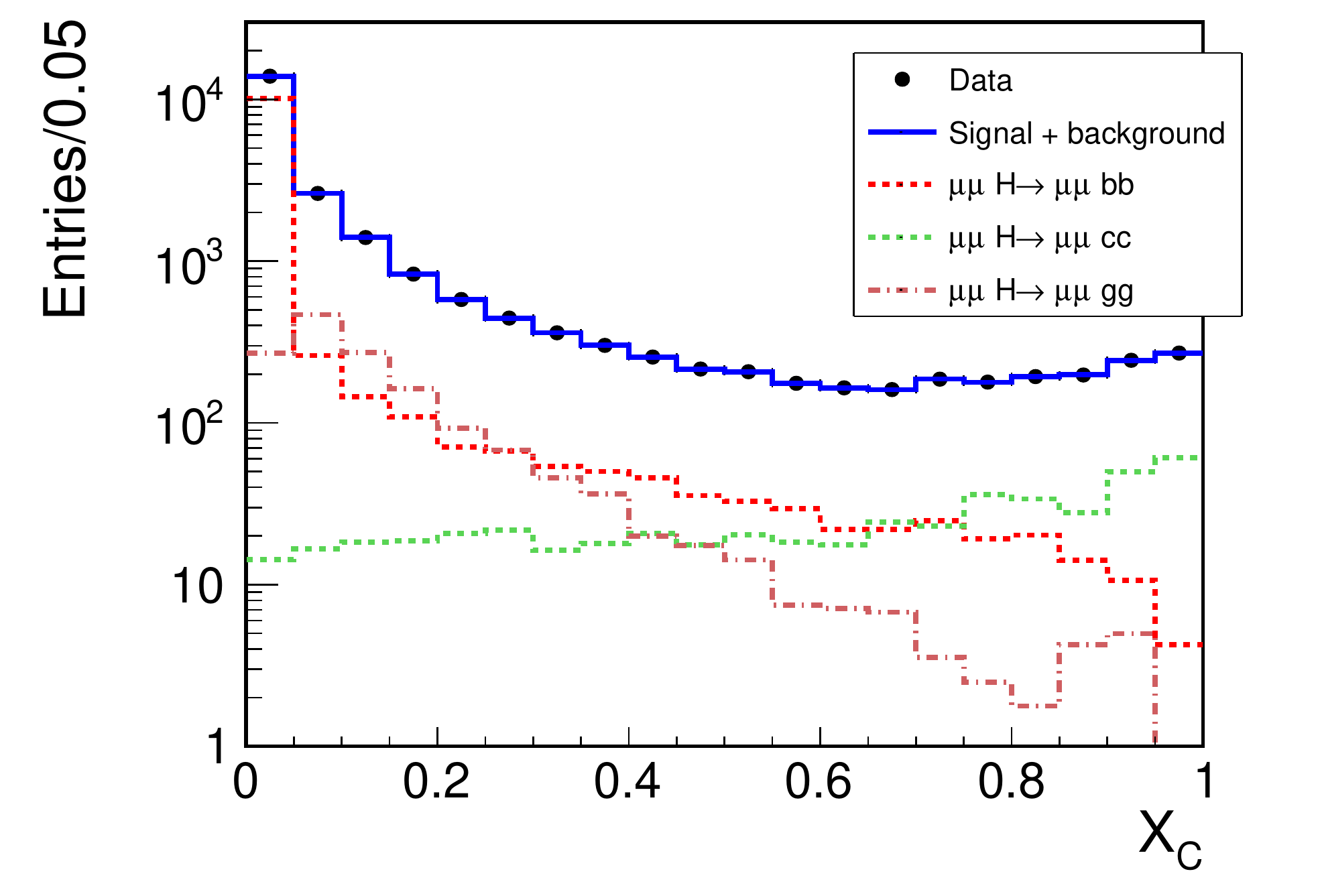}
    \end{minipage}
}
\subfigure[]
{
    \begin{minipage}[b]{0.47\textwidth}
    \includegraphics[width=\textwidth]{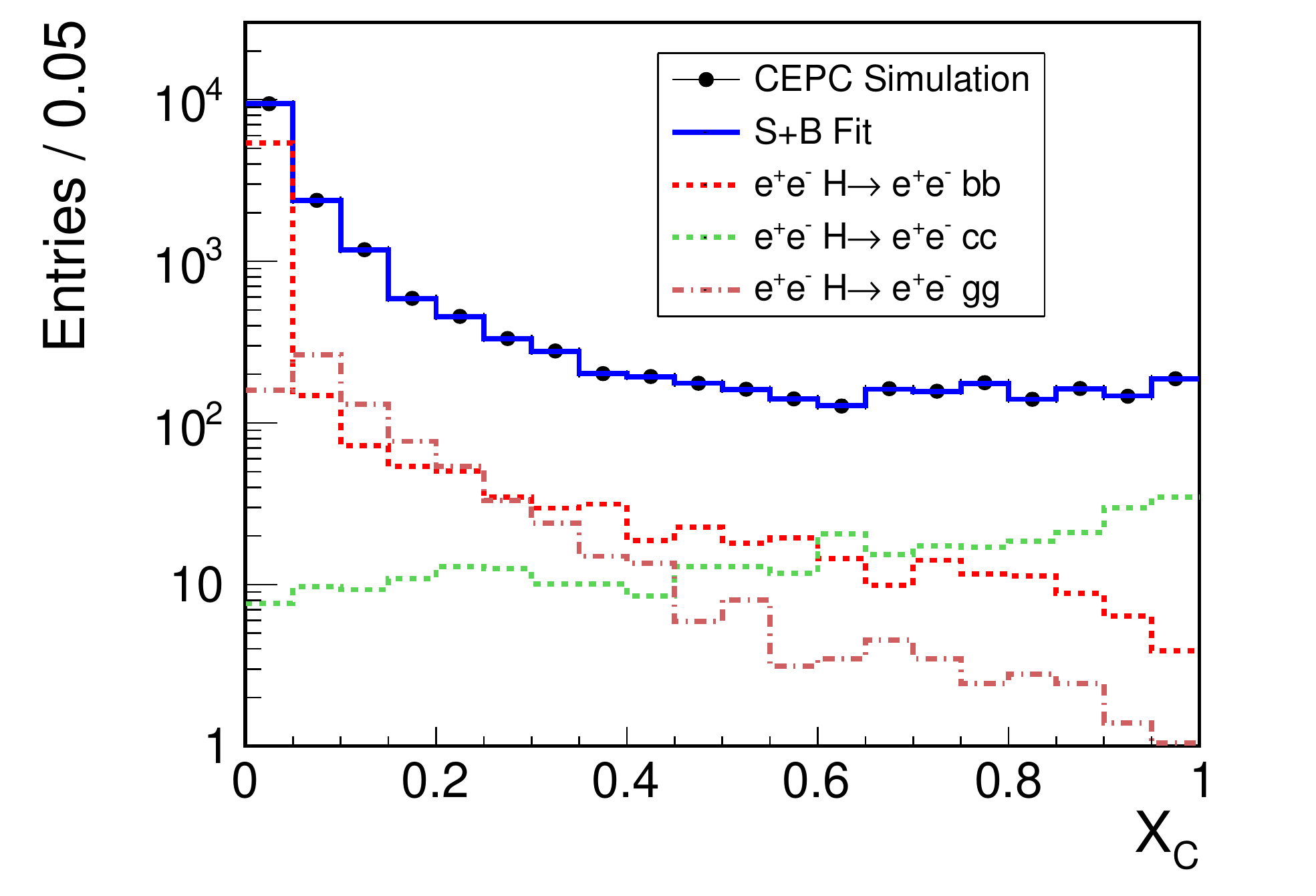}
    \end{minipage}
}
\figcaption{\label{fig:recoilmass_fit} 3D-fit result projected on three dimension: (a) fit result projected
on recoil mass distribution in $\mmh$ channel, (b) fit result projected
on recoil mass distribution in $\eeh$ channel, (c) fit result projected on B-likeness distribution in $\mmh$ channel, (d) fit result projected on B-likeness distribution in $\eeh$ channel, (e) fit result projected on C-likeness distribution in $\mmh$ channel, (f) fit result projected on C-likeness distribution in $\eeh$ channel.}
 \end{figure}

The fit program minimizes the negative logarithm of the likelihood function $-\sum_i w_i \log L(M_i,X^i_B,X^i_C)$, in which $L$ is the likelihood function presented in {Eq.(\ref{for:Fit_Likelihood})}; $w_i$, $X^i_B$, $X^i_C$ and $M_i$ are event weight\footnote{In this analysis the event weight are used for normalization, as mentioned in Sec.\ref{sec:MC_Sample}}, $X_B$, $X_C$ and $M^{l\bar{l}}_{\mathrm{recoil}}$ of event $i$ respectively. The summation is applied by including all the events in the fit region.
The fit results of the simulated data are shown in Fig.~\ref{fig:recoilmass_fit}, in which one can find that the model describes the data very well.
\par
 \end{multicols}
\begin{multicols}{2}

\section{Uncertainties of Measurements}
\label{sec:uncertainties}
\begin{figure}
\centering
\subfigure[]
{
   \begin{minipage}[b]{0.47\textwidth}
   \includegraphics[width=\textwidth]{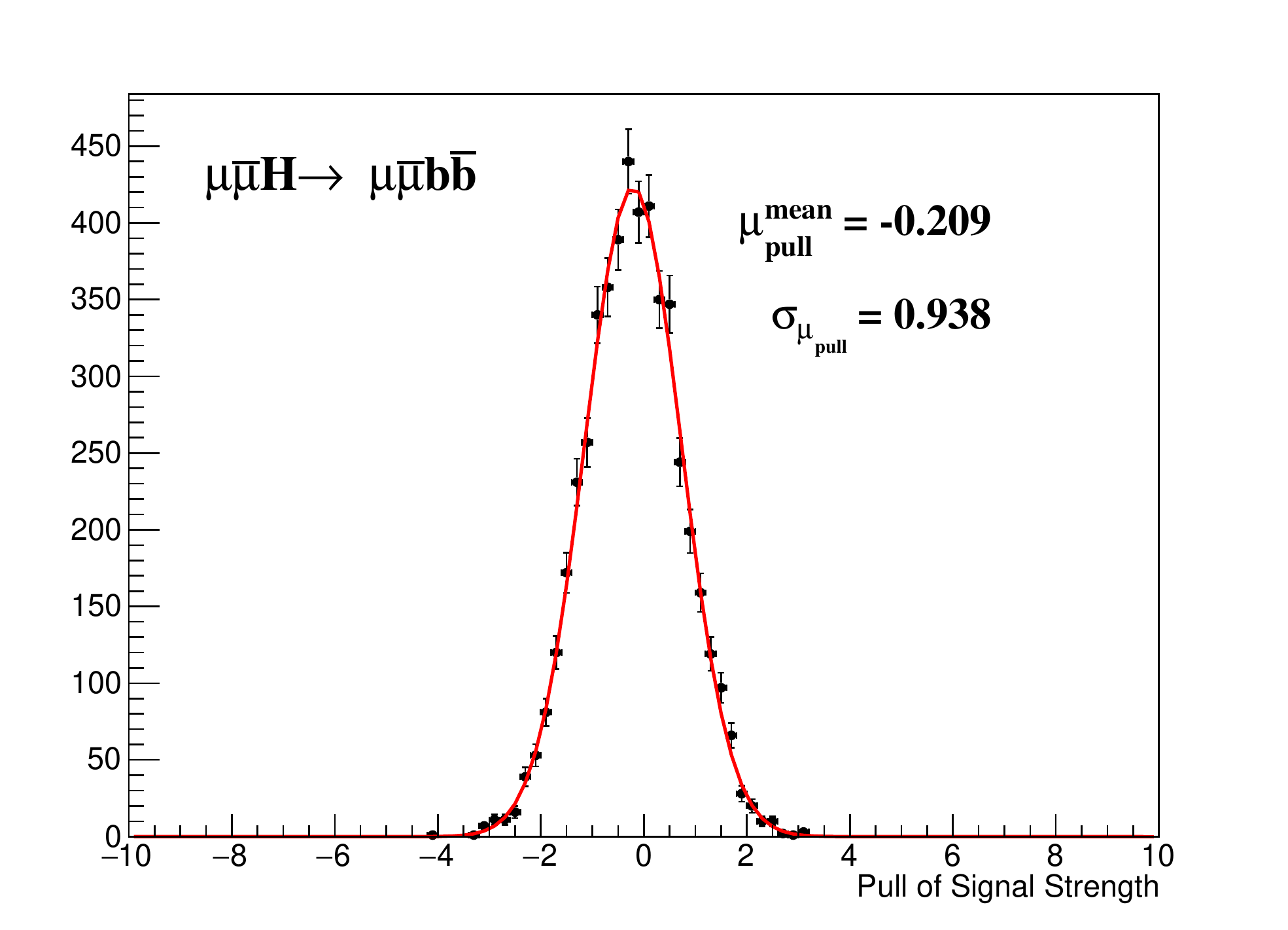}
   \end{minipage}
}
\subfigure[]
{
   \begin{minipage}[b]{0.47\textwidth}
   \includegraphics[width=\textwidth]{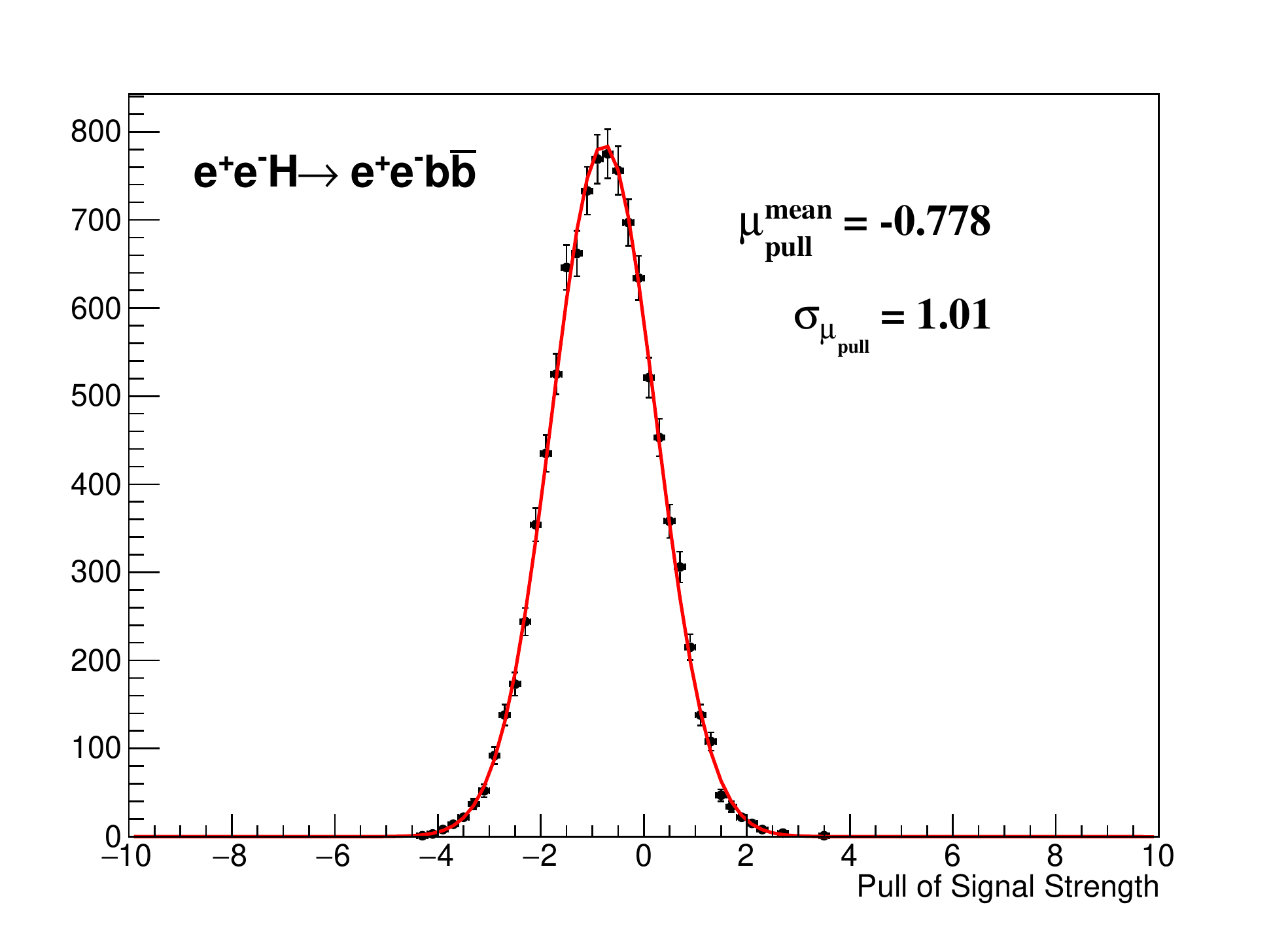}
   \end{minipage}
}
\subfigure[]
{
   \begin{minipage}[b]{0.47\textwidth}
   \includegraphics[width=\textwidth]{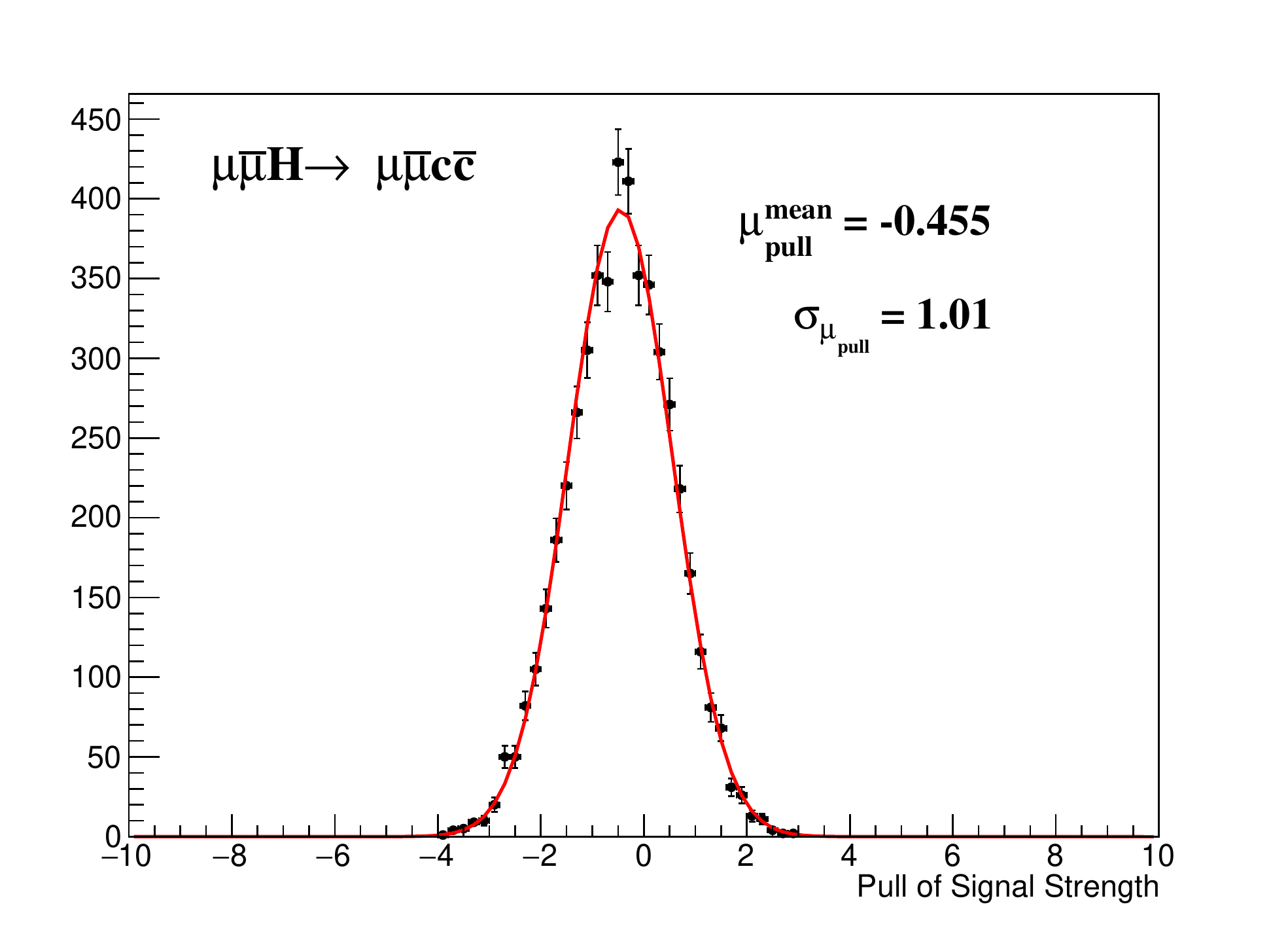}
   \end{minipage}
}
\subfigure[]
{
   \begin{minipage}[b]{0.47\textwidth}
   \includegraphics[width=\textwidth]{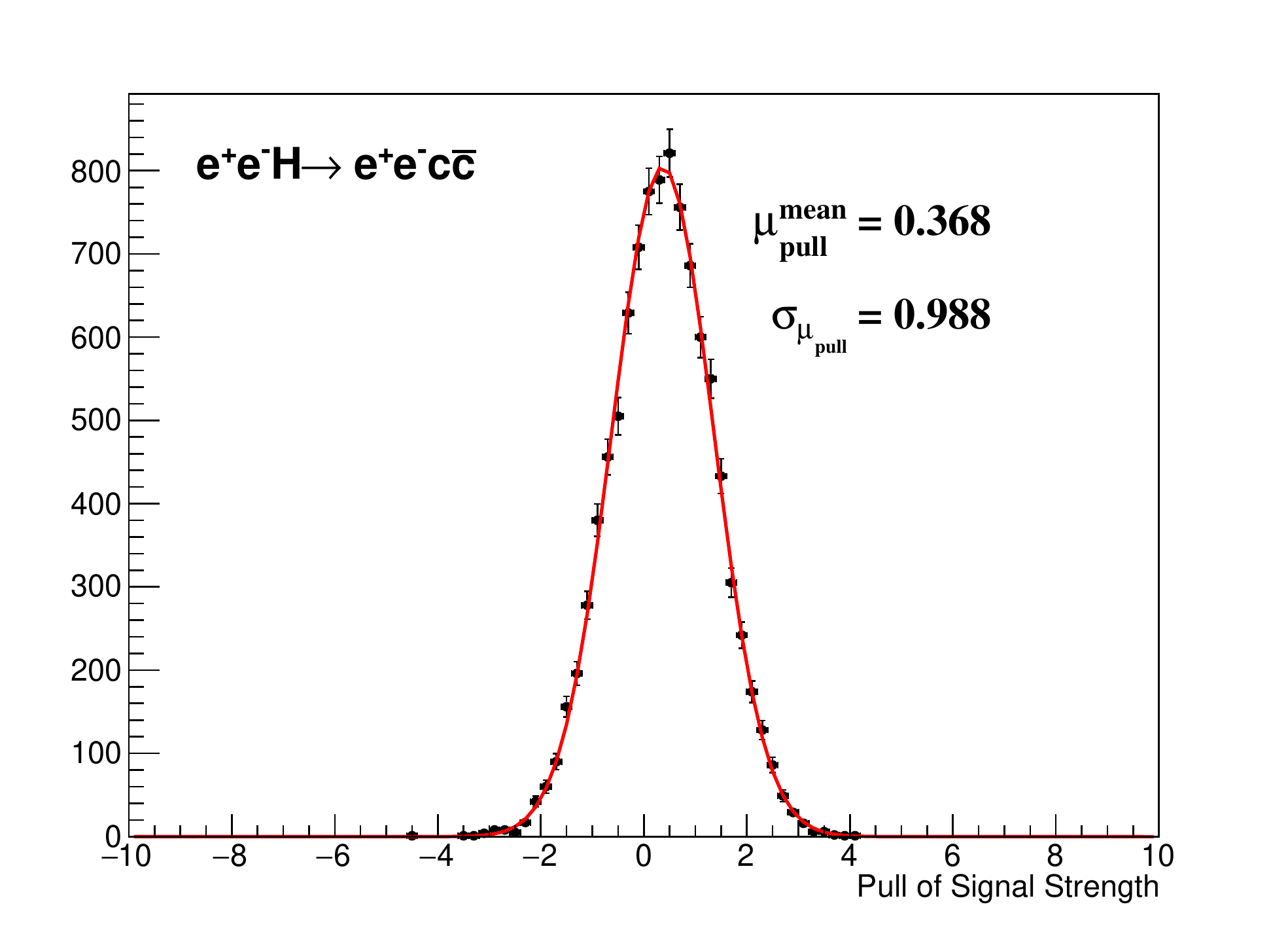}
   \end{minipage}
}
\subfigure[]
{
   \begin{minipage}[b]{0.47\textwidth}
   \includegraphics[width=\textwidth]{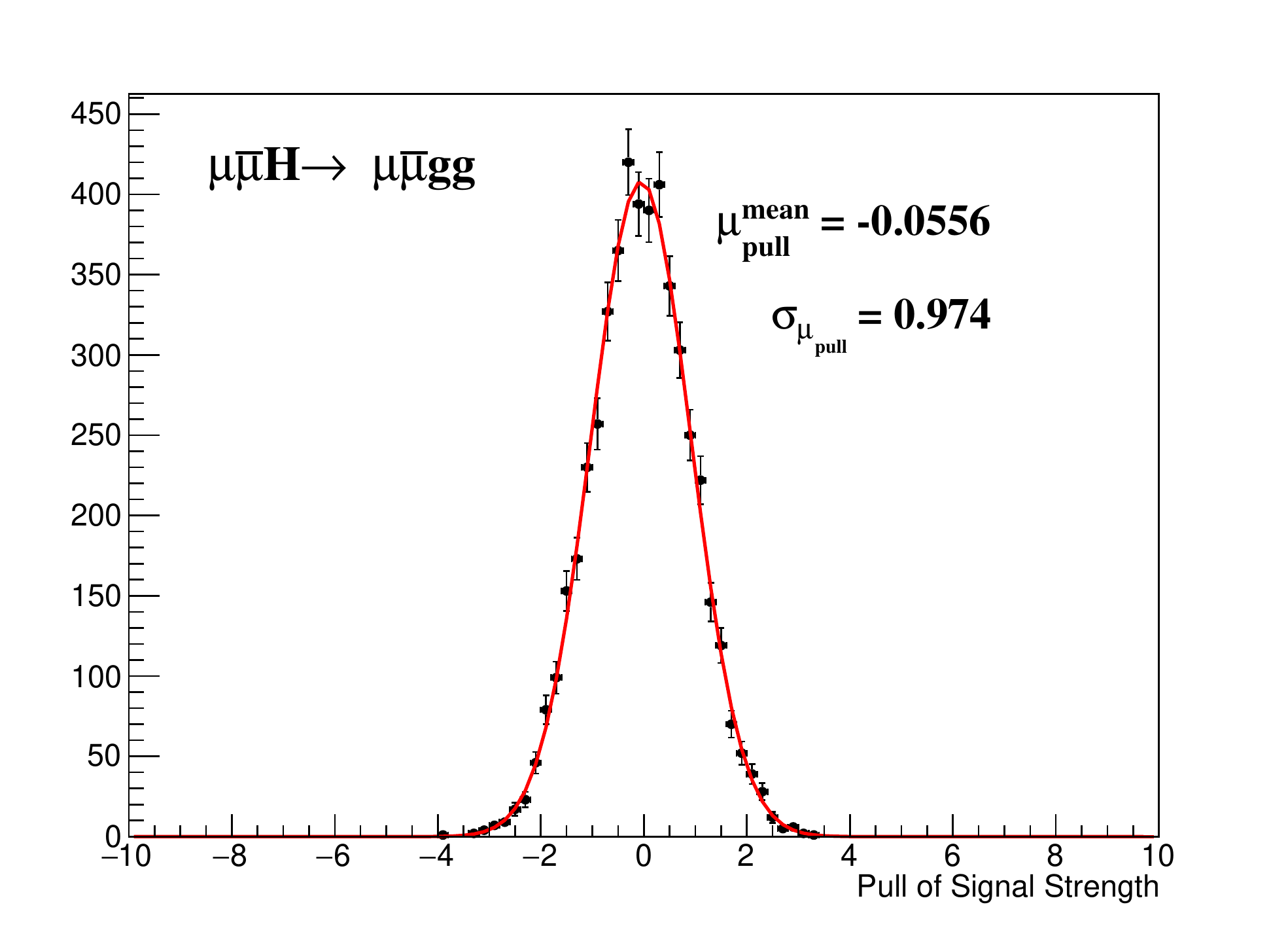}
   \end{minipage}
}
\subfigure[]
{
   \begin{minipage}[b]{0.47\textwidth}
   \includegraphics[width=\textwidth]{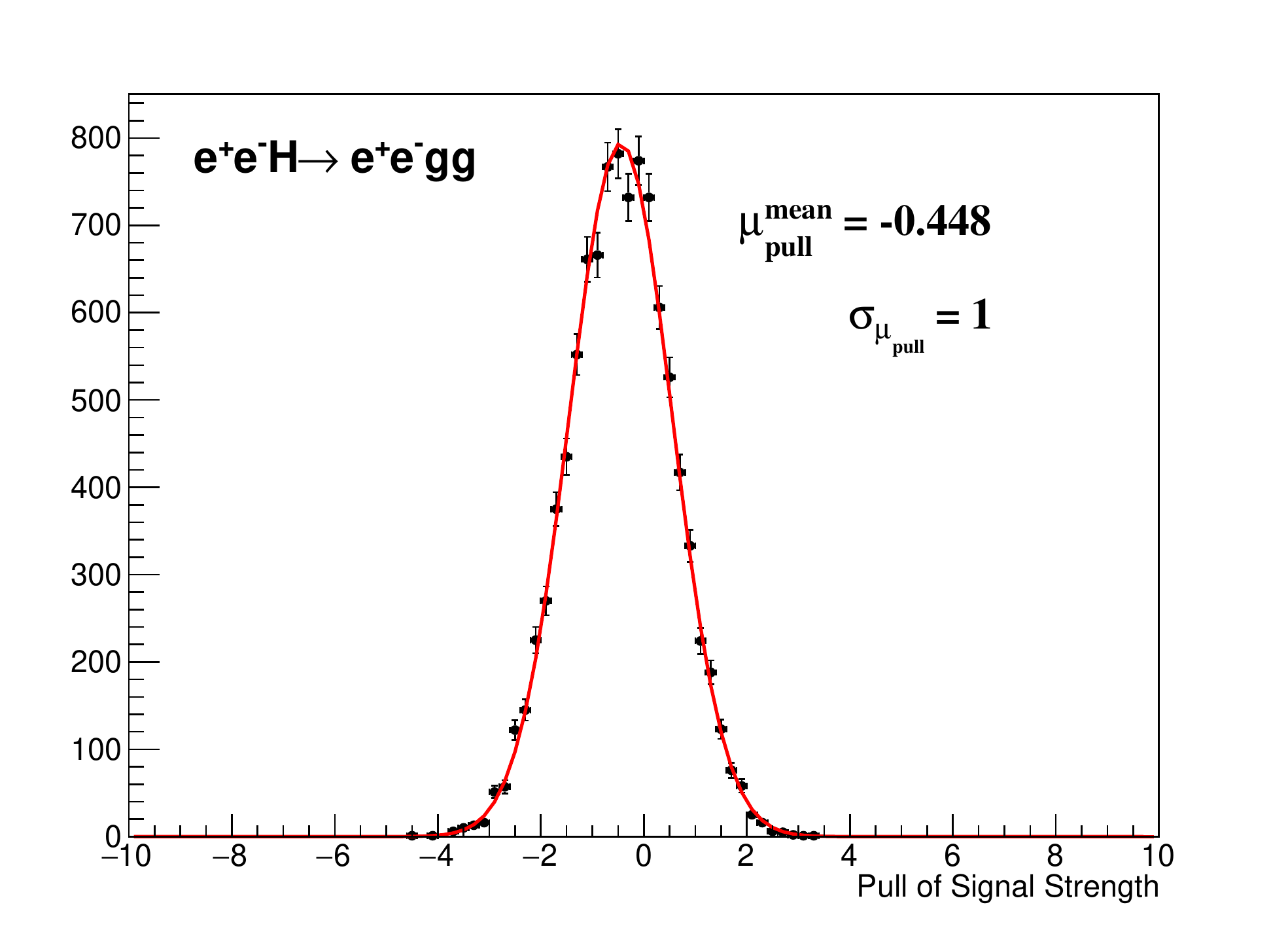}
   \end{minipage}
}
\figcaption{\label{fig:toyMC}Toy MC test result in terms of pull of signal strength in each channel. The pull distributions are fitted with Gaussian function.}
\end{figure}

The statistical uncertainty was estimated by using toyMC method that includes 6000~(\mmh~ channel) and 10000~(\eeh~ channel) iterations.
In each iteration, the 'data' is filled in a 3D histogram with dimension of $(M^{l\bar{l}}_{\mathrm{recoil}},X_B,X_C)$.
In each bin of the histogram, the event yields fluctuated according to a Poisson distribution.
Binned fit with the model described in {Sec.~\ref{sec:templatefit}} is then applied to the fluctuated histogram.
The statistical uncertainty of signal event yields can be estimated according to dispersion of fit results of the toyMC test.
The results of toyMC test for $H\to\bpair$, $H\to\cpair$ and $H\to gg$ are represented in Fig.~\ref{fig:toyMC}, in terms of pull of fitted signal events number for toyMC samples in the signal region which conform well to the standard normal distribution.
\par


The systematic uncertainties from luminosity, lepton identification and selection efficiencies, $Z\to\mupair/\elpair$ modeling and ISR correction factor~\cite{ISR} are included in the measurements of $\sigma^{H->\bpair/\cpair/\gpair}_{\llh}$.
However, these uncertainties are also included in the measurement of inclusive Higgs boson production cross section associated with lepton pair $\sigma_{l\bar{l}H}$, like that presented in Ref.~\cite{CEPC:recoilmass}.
To get the branching fractions of $H\to \bpair/\cpair/\gpair$, one need to divide the measured $\sigma^{H->\bpair/\cpair/\gpair}_{\llh}$ by $\sigma_{l\bar{l}H}$. As a consequence the systematic uncertainties discussed above will be canceled.
So here we can ignore these uncertainties.
\par
The fit method described in {Sec.\ref{sec:templatefit}} has two types of systematic uncertainty. The first kind is due to imperfect modeling of the PDF in the likelihood in Eq.~(\ref{for:Fit_Likelihood}).
The inaccurate modeling of $M^{l\bar{l}}_{\mathrm{recoil}}$ distribution and the bias in the prediction of $F_{\mathrm{flavor}}(X_B,X_C)$ can lead to such kind of uncertainty.
The latter will be discussed later and here we only focus on the recoil mass modeling.
The recoil mass functions for signal and background are verified by fitting the signal and background datasets alone respectively.
These results demonstrate that the function describes the shape very well. The shape parameters are left to float in the fit, except for those of the tail function in signal recoil mass.
These tail function can be studied by comparing the track momentum and energy resolution in data and MC. So far we assume the resolution is well simulated in MC sample.
\par
The other kind of systematic uncertainty in the fit comes from the uncertainty of fixed parameters,
including the event yields of $\eeh$ or $\mmh$ background, as well as the yields of all irreducible background.
We conservatively set $H\to WW^*$ and $H\to ZZ^*$ event yields 5\% higher and lower than the MC prediction,
and vary the yields of non-$ZZ$ backgrounds by $\pm100$\%, to estimate the corresponding systematic uncertainty.
The systematic uncertainties discussed above are included in the row of 'Fixed Background' in Tab.~\ref{tab:uncertainties}.
\par
In signal events extra leptons~(leptons other than the 2 primary leptons) comes from leptonic decay of heavy flavor quarks.
Systematic uncertainty of extra isolated lepton veto efficiency for $H\to\bpair$ and $H\to\cpair$ can be estimated using the $\bpair$ and $\cpair$ events produced at $Z$-pole respectively.
By selecting the two-jets events, requiring both of the two jets tagged as $b$-jets and looking for an isolated lepton, the efficiency of lepton veto can be studied 
with 2 billion \bpair~events in the $Z$-pole sample.
The veto efficiency can be studied in a precision so high that it has no visible impact to the analysis presented, due to the high statistics of $Z-$pole sample and very low rate of occurrence of extra leptons.
The impact to $H\to\cpair$ can be studied in similar way. By requiring both jets to be $c$-tagged in $Z-$pole sample and looking for isolated leptons, one can also study the impact with a precision high enough to have any visible impact to the analysis.
For $H\to\gpair$ we assume the MC have $\pm50$\% of uncertainty in predicting the lepton-veto efficiency, and estimate the impact to be negligible in this analysis.\par
The jets' particle multiplicity, jet angular distribution and $Y_k$ value can be studied in very high precision with high statistics $Z-$pole data. Correspondingly, the systematic uncertainties of the efficiencies of jets' particle multiplicity cut, jet $\cos \theta$ cut and $Y_k$ are negligible.\par
The systematic uncertainty of the efficiency of jet pair invariant mass cut can be
estimated from the jet energy resolution. We apply a smearing on jet pair mass distribution according to a gaussian distribution corresponding to the jet energy resolution.
We take the value of 4\% as the jet energy resolution from the CEPC pre-CDR~\cite{CEPC_preCDR} and calculate the uncertainty on the event yields of $H\to \bpair$, $H\to\cpair$ and $H\to\gpair$ are $^{+0.68\%}_{-0.20\%}$, $^{+0.43\%}_{-1.08\%}$ and $^{+0.71\%}_{-1.68\%}$ respectively. The uncertainties of extra lepton veto, the jet angular and reconstructed particles multiplicity and jet pair mass resolution are included in the row of 'Event Selection' in Tab.~\ref{tab:uncertainties}.\par
Since flavor tagging method is implemented via flavor template fit, the flavor tagging systematic uncertainty is directly caused by the difference between the templates from the MC prediction and the templates in data.
Evaluating such differences demands delicate flavor tagging commissioning and calibration.
Although no such commissioning or calibration has been done yet, we can estimate the systematic uncertainty by assuming a difference between data and MC after the calibration was applied, and this difference is subsequently studied in terms of its impact on the $H\to\bpair/\cpair/\gpair$ branching fractions measurement.
We select $ZZ\to \qpair+\mupair$ events as a control sample, which has a purity of 99.6\%, and assuming a data-MC comparison has been done on the template distribution on this control sample.
The estimation of the data-MC agreement is limited by the statistic uncertainty of the control sample, and the knowledge of the flavor components of hadronic $Z-$decay.
For example, more than 80\% of the $\bpair$ events is concentrated in the region with highest $b$-likeness($b$-likeness$>$0.95) and lowest $c$-likeness($c$-likeness$<$0.05).
If, due to some kinds of $b-$tagging systematic uncertainty, the $Z\to\bpair$ events fraction in this most concentrate bin changed, the data-MC disagreement would be increase.
There are $1.92\times 10^4$ ~$Z\to\bpair$ in this bin, which has a statistic uncertainty 0.72\%. The current combined measurements of $R_b$, defined as $Br(Z\to\bpair)/Br(Z\to\qpair)$, has the uncertainty of 0.31\%~\cite{pdg_Z}.
So the data and MC can be compared in the precision of $\sqrt{0.72\%^2+0.31\%^2}=0.78\%$. Scaling the contents in this bin up and down by 0.78\% in the $\bpair$ template, one can estimate the uncertainty to $H\to\bpair$, $H\to\cpair$ and $H\to\gpair$ are $^{-0.4\%}_{+0.2\%}$, $^{+3.7\%}_{-5.0\%}$ and $^{+0.2\%}_{-0.7\%}$ respectively.
Had we use a much larger control sample (hadronic events at the $Z$-pole), and had a better understanding on the relationship between the flavor tagging variables and kinematic feature, the uncertainty will be further reduced.\par
The statistical uncertainty and systematic uncertainty discussed in this section are summarized in Tab.~\ref{tab:uncertainties}.\par
\end{multicols}

\begin{center}
\tabcaption{ \label{tab:uncertainties}Uncertainties on $\sigma^{\bpair}_{\llh}$, $\sigma^{\cpair}_{\llh}$ and $\sigma^{\gpair}_{\llh}$.}
\footnotesize
\begin{tabular}{c|c|c|c|c|c|c}\hline
      Higgs boson production       & \multicolumn{3}{c|}{\mmh}           &  \multicolumn{3}{c}{\eeh} \\ \hline
      Higgs boson decay              &  $H\to \bpair$ &   $H\to\cpair$    &  $H\to\gpair$
             &  $H\to \bpair$ &   $H\to\cpair$    &  $H\to\gpair$    \\ \hline
  Statistic Uncertainty   &      1.1\%           &    10.5\%         &     5.4\%
       &     1.6\%           &    14.7\%         &     10.5\%   \\ \hline

 Fixed Background
             & \tabincell{c}{-0.2\% \\ +0.1\%}
                              &    \tabincell{c}{+4.1\% \\ -4.2\%}
                                                  &  7.6\%
             &  \tabincell{c}{-0.2\% \\ +0.1\%}
                              &    \tabincell{c}{+4.1\% \\ -4.2\%}
                                                  &  7.6\%  \\ \hline
Event Selection
             &  \tabincell{c}{+0.7\% \\ -0.2\%}
                              &  \tabincell{c}{+0.4\% \\ -1.1\%}
                                                  &   \tabincell{c}{+0.7\% \\ -1.7\%}
             &  \tabincell{c}{+0.7\% \\ -0.2\%}
                              &  \tabincell{c}{+0.4\% \\ -1.1\%}
                                                  &   \tabincell{c}{+0.7\% \\ -1.7\%}
                                                    \\ \hline
Flavor Tagging
             &  \tabincell{c}{-0.4\% \\ +0.2\%}
                              &  \tabincell{c}{+3.7\% \\ -5.0\%}
                                                  &   \tabincell{c}{+0.2\% \\ -0.7\%}
             &  \tabincell{c}{-0.4\% \\ +0.2\%}
                              &  \tabincell{c}{+3.7\% \\ -5.0\%}
                                                  &   \tabincell{c}{+0.2\% \\ -0.7\%}
                                                    \\ \hline
  Combined Systematic Uncertainty  &     \tabincell{c}{+0.7\% \\ -0.5\%}
                              &  \tabincell{c}{+5.5\% \\ -6.6\%}
                                                  &       \tabincell{c}{+7.6\% \\ -7.8\%}
              &     \tabincell{c}{+0.7\% \\ -0.5\%}
                              &  \tabincell{c}{+5.5\% \\ -6.6\%}
                                                  &       \tabincell{c}{+7.6\% \\ -7.8\%}          \\ \hline

\bottomrule
\end{tabular}
\end{center}
\begin{multicols}{2}
\section{Conclusion}\label{sec:summary}
The measurements of branching fractions of $H\to\bpair/\cpair/\gpair$ are studied in $\mmh$ and $\eeh$ process, in the scenario of analysing 5000~\ifb~$\elpair$ collision data with $\sqrt{s}$ of 250~\GeV~in CEPC.
The statistical uncertainty on $\sigma^{\bpair}_{\llh}$, $\sigma^{\cpair}_{\llh}$ and $\sigma^{\gpair}_{\llh}$ measurements are estimated around 1.1\%,  10.5\% and 5.4\% respectively in $\mmh$ channel, and 1.6\%, 14.7\% and 10.5\% respectively in $\eeh$ channel.
The systematic uncertainties on the branching fraction measurements are also studied, which are around 0.6\%, 6\% and 8\% for $\bpair$, $\cpair$ and $\gpair$ final states respectively.
The high precision of this measurement benefits from the distinct signature of events with the Higgs boson and clean background in electron-positron collider, as well as the model independent analysis.
As a comparison, by combining the extrapolated results in ATLAS and CMS in the scenario of the High Luminosity LHC(HL-LHC)~\cite{HL_Yellow_Book}, the $H\to\bpair$ branching fraction precision is expected to be 4.4\%, in which the statistic uncertainty, the systematic uncertainty and theoretical uncertainty are 1.5\%, 1.3\% and 4.0\% respectively.
This study demonstrates the feasibility of precise measurement of Higgs
Yukawa coupling to quarks at the CEPC.\par

\section{Acknowledgment*}
We would like to thank the CEPC higgs physics working group for the valuable discussions as well as the software and physics work without which this work couldn't have been accomplished.\par

\end{multicols}

\vspace{-1mm}
\centerline{\rule{80mm}{0.1pt}}
\vspace{2mm}

\begin{multicols}{2}

\end{multicols}

\clearpage

\end{CJK*}

\begin{thebibliography}{90}

\vspace{3mm}

%
%
\bibitem{Higgs_CMS}CMS Collaboration, Phys. Lett. B, {\bf 716}: 30~(2012)
\bibitem{Higgs_ATLAS}ATLAS Collaboration, Phys. Lett. B, {\bf 716}: 1~(2012)
\bibitem{BEH} P.W.Higgs, Phys. Rev. Lett, {\bf 13}: 508(1964)
\bibitem{BEH2} F.Englert and R.Brout, Phys. Rev. Lett, {\bf 13}: 321~(1964)
\bibitem{BEH3} G.S.Guralnik, C.R.Hagen and T.W.B.Kibble, Phys. Rev. Lett, {\bf 13}: 585~(1964)
\bibitem{Higgs_qq_1} S. Gorishnii et al., Mod. Phys. Lett. A, {\bf 5}: 2703~(1990);
                 S. Gorishnii et al., Phys. Rev. D, {\bf 43}: 1633~(1991);
                 A.L. Kataev and V.T. Kim, Mod. Phys. Lett. A, {\bf 9}: 1309~(1994);
                 L.R. Surguladze, Phys. Lett. B, {\bf 341}: 60~(1994);
                 S. Larin, T. van Ritbergen, and J. Vermaseren, Phys. Lett. B, {\bf 362}: 134~(1995); K. Chetyrkin and A. Kwiatkowski, Nucl. Phys. B, {\bf 461}: 3~(1996);
                 K. Chetyrkin, Phys. Lett. B, {\bf 390}: 309~(1997);
                 P.A. Baikov, K.G. Chetyrkin, and J.H. Kuhn, Phys. Rev. Lett: {\bf 96}, 012003~(2006)
\bibitem{Higgs_qq_2} J. Fleischer and F. Jegerlehner, Phys. Rev. D, {\bf 23}: 2001~(1981);
                 D. Bardin, B. Vilensky, and P. Khristova, Sov. J. Nucl. Phys,{\bf 53}: 152~(1991);
                 A. Dabelstein and W. Hollik, Z. Phys. C, {\bf 53}: 507~(1992);
B.A. Kniehl, Nucl. Phys. B, {\bf 376}: 3~(1992);
\bibitem{Higgs_gg_1} T. Inami, T. Kubota, and Y. Okada, Z. Phys. C, {\bf 18}: 69~(1983);
                 K.G. Chetyrkin, B.A. Kniehl, and M. Steinhauser, Phys. Rev. Lett, {\bf 79}: 353~(1997);
                 P.A. Baikov and K.G. Chetyrkin, Phys. Rev. Lett, {\bf 97}: 061803~(2006)
\bibitem{Higgs_gg_2} M. Spira et al., Nucl.Phys.B, {\bf 453}: 17~(1995)
\bibitem{VH_bb_atlas}ATLAS Collaboration, JHEP,{\bf 01}: 069~(2015)
\bibitem{VH_bb_cms}CMS Collaboration, Phys.Rev.D, {\bf 89}: 012003~(2014)
\bibitem{ttH_bb_cms_1} CMS Collaboration, JHEP, {\bf 05}: 145~(2013)
\bibitem{ttH_bb_cms_2} CMS Collaboration, JHEP, {\bf 09}: 087~(2014)
\bibitem{ttH_bb_atlas_1} ATLAS Collaboration, Eur.Phys.J.C, {\bf 75}:349~(2015)
\bibitem{VBF_bb_atlas}ATLAS Collaboration, JHEP, {\bf 11}: 112~(2016)
\bibitem{VBF_bb_cms}CMS Collaboration, Phys.Rev.D, {\bf 92}: 032008~(2015)
\bibitem{Tevatron_VH_bb}CDF and D0 Collaboration, Phys.Rev.Lett, {\bf 109}, 071804(2012)
\bibitem{higgs_atlas_cms_combine}CMS Collaboration and ATLAS Collaboration, JHEP, {\bf 08}: 045~(2016)
\bibitem{atlas_vhbb_new}ATLAS Collaboration, Phys. Lett. B, {\bf 786}: 59~(2018)
\bibitem{cms_vhbb_new}CMS Collaboration, Phys. Rev. Lett, {\bf 121}: 121801~(2018)
\bibitem{H_cc_atlas}ATLAS Collaboration, Phys.Rev.Lett, {\bf 120}:211802~(2018)
\bibitem{ILC} H.Baer, T.Barklow, K.Fujii, Y.Gao, A.Hoang, $et al.$, \emph{The International Linear Collider Technical Design Report - Volume 2: Physics}, arxiv:1306.6352
\bibitem{TLEP} A.Abada $et al.$, Eur.Phys.J.ST {\bf 228}, no. 2: 261-623~(2019)
\bibitem{CLIC} The CLIC, CLICdp Collaborations, \emph{Updated baseline for a staged Compact Linear Collider}, arxiv:1608.07537
\bibitem{CEPC_preCDR}M.Ahmad and others, The CEPC-SPPC Study Group \emph{CEPC-SPPC Preliminary Conceptual Design Report, Vol I}, IHEP-EP-2015-01, 201
\bibitem{Higgs_strahlung_1} B.W. Lee, C. Quigg, and H.B. Thacker, Phys. Rev. D, {\bf 16}: 1519~(1977)
\bibitem{Higgs_strahlung_2} J. Ellis, M.K. Gaillard, and D.V. Nanopoulos, Nucl. Phys. B, {\bf 106}, 292~(1976)
\bibitem{Higgs_strahlung_3} B.L. Ioffe and V.A. Khoze, Sov. J. Nucl. Phys, {\bf 9}: 50~(1978)
\bibitem{WW_fusion} D.R.T. Jones and S. Petcov, Phys. Lett. B, {\bf 84}: 440~(1979);
                    R.N. Cahn and S. Dawson, Phys. Lett. B, {\bf 136}: 196~(1984);
                    G.L. Kane, W.W. Repko, and W.B. Rolnick, Phys. Lett. B, {\bf 148}: 367~(1984);
                    G. Altarelli, B. Mele, and F. Pitolli, Nucl. Phys. B, {\bf 287}: 205~(1987);
                    W. Kilian, M. Kramer, and P.M. Zerwas, Phys. Lett. B, {\bf 373}: 135~(1996)
\bibitem{ILC:Hbbccgg}Hiroaki Ono and Akiya Miyamoto, Eur.Phys.J.C, {\bf 73}: 2343~(2013)
\bibitem{CEPC_Higgs} Fen-fen An $et al.$, Chin.Phys.C, {\bf 43}(4): 043002~(2019)
\bibitem{PFA}J.Brient, Proceedings of 11th International Conference of Calorimetry in High Energy Physics, 2004, 445--451
\bibitem{CEPC_CDR} The CEPC Study Group, \emph{CEPC Conceptual Design Report-Vol II}, IHEP-CEPC-DR-2018-02
\bibitem{Wizard_1} W.Kilian, T.Ohl, J. Reuter, Eur.Phys.J.C, {\bf 71}: 1742~(2011)
\bibitem{PYTHIA64} T.Sj$\mathrm{\ddot{o}}$strand, S.Mrenna and P.Skands, JHEP, {\bf 05}: 026~(2006)
\bibitem{CEPC_Generation} Xin Mo, Gang Li, Man-Qi Ruan, and Xin-Chou Lou, Chin. Phys. C, {\bf 40}(3): 033001~(2016)
\bibitem{mokka}P.Mora de Freitas and H.Videau, LC-TOOL-2003-010, 2003:623--627
\bibitem{Geant4}GEANT4 Collaboration, Nucl.Instrum.Meth, {\bf 506}: 250-303~(2003)
\bibitem{ArborPFA_1}CMS Collaboration, CMS Particle Analysis Summary, CMS-PAS-PFT-09-001~(2009)
\bibitem{ArborPFA_2}CMS Collaboration, CMS Particle Analysis Summary, CMS-PAS-PFT-10-001~(2010)
\bibitem{CEPC:reconstruction} Man-Qi Ruan $et al.$, Eur.Phys.J.C, {\bf 78}(5), 426~(2018)
\bibitem{LICH}Dan Yu, Man-Qi Ruan, V. Boundry and H Videau, Eur.Phys.J.C, {\bf 77}, 591~(2017)
\bibitem{BDT} Byron P. Roe, Hai-Jun Yang, Ji Zhu, Yong Liu, I. Stancu and G. McGregor, J.NIMA, {\bf 543}:577--584~(2005)
\bibitem{LCFIPlus}T. Suehara and T. Tanabe, J.NIMA, {\bf 808}: 109-116~(2015)
\bibitem{Durham}S.Catani, Y.L.Dokshitzer, M.Olsson, G.Turnock, B.Webber, Phys.Lett.B, {\bf 269}: 432-438~(1991)
\bibitem{ISR} M. Greco, G. Montagna, O.Nicrosini, F. Piccinini and G.Volpi, Phys.Lett.B, {\bf 777}:294-297~(2018)
\bibitem{CEPC:recoilmass}Zhen-Xing Chen, Ying Yang, Man-Qi Ruan, Da-Yong Wang, Gang Li, Shan Jin, Yong Ban, Chin. Phys. C, {\bf 41}(2): 023003~(2017)
\bibitem{pdg_Z} M.Tanabashi $et al.$, (Particle Data Group), Phys.Rev.D, {\bf 98}: 030001~(2018)
\bibitem{HL_Yellow_Book} R.Abdual Khalek $et al.$, \emph{Higgs Physics at the HL-LHC and HE-LHC}, CERN-LPCC-2018-04~(2109), arxiv: 1902.00134
\end{thebibliography}
\end{document}